\documentclass[pre,showpacs,two column,eqsecnum]{revtex4}

\usepackage{amsfonts}
\usepackage{amssymb}
\usepackage{mathtools}
\usepackage{graphicx}

\begin{document}

\preprint{\bibliographystyle{revtex4}}

\title{Work, work fluctuations, and the work distribution in a thermal non-equilibrium steady state}

\author{T. R. Kirkpatrick, J.R. Dorfman, and J.V. Sengers}
\affiliation{Institute for Physical Science and Technology, University of Maryland, College Park, Maryland 20742, USA}

\date{\today}

\begin{abstract}
Long-ranged correlations generically exist in non-equilibrium fluid systems. In the case of a non-equilibrium steady state caused by a temperature gradient the correlations are especially long-ranged and strong. The anomalous light scattering predicted to exist in these systems is well-confirmed by numerous experiments. Recently the Casimir force or pressure due to these fluctuations or correlations have been discussed in great detail. In this paper the notion of a Casimir work is introduced and a novel way to measure the non-equilibrium Casimir force is suggested. In particular, the non-equilibrium Casimir force is related to a non-equilibrium heat and not, as in equilibrium, to a volume derivative of an average energy. The non-equilibrium work fluctuations are determined and shown to be very anomalous  compared to equilibrium work fluctuations. The non-equilibrium  work distribution is also computed, and contrasted to work distributions in systems with short-range correlations. Again, there is a striking difference in the two cases. Formal theories of work and work distributions in non-equilibrium steady states are not explicit enough to illustrate any of these interesting features.

\end{abstract}


\maketitle

\section{INTRODUCTION}
\label{sec:I}

The study of fluctuations in non-equilibrium (NE) has a long history. For reviews see \cite{Dorfman_Kirkpatrick_Sengers_1994,Schmittmann_Zia_1995, Belitz_Kirkpatrick_Vojta_2005, DeZarate_Sengers_2006,Derrida_2007, Sengers_DeZarate_Kirkpatrick_2015}. The most interesting aspect of the correlations in a non-equilibrium steady state (NESS) is that in general they are of long range as long as the system has conserved, hydrodynamic, variables or other soft modes.

The two general systems where the effects are the largest, are a simple fluid in a constant temperature gradient and a binary mixture with either a concentration gradient or a temperature gradient. For example, in the early $1980$'s
it was predicted \cite{Kirkpatrick_Cohen_Dorfman_1982A,Kirkpatrick_Cohen_Dorfman_1982B}
that the temperature and density correlations in a simple fluid in a NESS
with a temperature gradient were extraordinarily long-ranged (in a
sense growing with system size). This prediction was subsequently
confirmed with great
precision in numerous small angle light scattering and shadowgraph experiments \cite{Law_et_al_1990,DeZarate_Sengers_2006}. For very small
angle scattering, the scattering was found to be larger than the equilibrium
scattering by a factor of $10^{5}$. All of this implies that the
statistics of fluctuations in non-equilibrium fluids will in general
be very different from those for fluctuations in the same fluid in
an equilibrium state \cite{Kirkpatrick_Dorfman_2015}.

In the $1990$'s, different aspects of fluctuations in fluids maintained in non-equilibrium
steady states, and in non-equilibrium fluids in general were considered. Of particular interest are
the studies of Evans, Cohen and Morris (ECM) \cite{Evans_Cohen_Morris_1993},
Evans and Searles \cite{Evans_Searles_1994}, and of Gallavotti and
Cohen \cite{Gallavotti_Cohen_1995}. 
In this earlier work, a focus was on non-equilibrium currents
and entropy production. For example, if $P_{xy\tau}$ is the time
average of the microscopic stress tensor, $P_{xy}$, over a time interval
$\tau$, then in a NESS with a steady shear rate, ECM studied
the probability distribution $\mathcal{P}(P_{xy\tau})$ in a $N-$particle
system. Interestingly, it depended exponentially on the entropy production, or irreversible work. This in turn has subsequently \cite{Evans_Searles_1995} been related to the $Gamma$-space NE distribution approach that was developed long ago \cite{MacLennan_1961, Zubarev_1962, Zubarev_1974}. See also \cite{Maes_Netocny_2010, Sasa_2014}. These NE distribution functions were used \cite{Kirkpatrick_Cohen_Dorfman_1982A, Kirkpatrick_Cohen_Dorfman_1982B} to show that non-equilibrium correlations are generically of long range.

Later on, Jarzynski \cite{Jarzynski_1997} derived a closely related identity that is now called the Jarzynski fluctuation theorem (JFT), 
\begin{equation}
\langle e^{-\beta W}\rangle=e^{-\beta\Delta F},
\end{equation}
where $\beta=1/(k_{B}T)$, $W$ is the total work, reversible plus irreversible, and $\Delta F$ is the free energy change. Here the angular brackets denote an average over a work distribution from one thermodynamic state to another, with $\Delta F$ the free energy differences between the two states. Because of poor numerical convergence properties \cite{Rodriguez-Gomez_et_al_2004, Daura_et_al_2010, Kim_et_al_2012, Talkner_et_al_2013}, this relation is not as useful in determining free energy difference as originally thought. Indeed, in Appendix A we show for a simple model with short-range correlations (an ideal gas), and argue more generally for systems with short-range correlations, that the fluctuations of $\Omega=e^{-\beta W}$ diverge exponentially as the system size increases. That is,
\begin{equation}
\epsilon_{\Omega}=\frac {\langle{\Omega}^2 \rangle -{\langle\Omega \rangle}^{2}}{{\langle\Omega \rangle}^{2}},
\end{equation}
is exponentially large in the system size. As a consequence of this, the JFT is generally only useful when applied to small systems \cite{Hummer_Szabo_2001, Liphardt_et_al_2002}, if the correlations are of short-range. To probe and understand the work distribution for macroscopic systems, naive considerations would suggest it is better to consider the work itself, since equilibrium statistical mechanics gives that the relative root mean-squared fluctuations of the pressure vanish as $O(1/\sqrt V)$, as the volume $V$ of the system increases.

As noted above, work is related to pressure. Very recently the average Casimir force or pressure due to the long-range correlations or fluctuations that exist in a NESS
have been discussed in great detail \cite{Kirkpatrick_DeZarate_Sengers_2013, Kirkpatrick_DeZarate_Sengers_2014,Kirkpatrick_DeZarate_Sengers_2015,Kirkpatrick_DeZarate_Sengers_2016a,Kirkpatrick_DeZarate_Sengers_2016b}\cite{Aminov_Kafri_Kardar_2015}. These NE fluctuation induced forces are generally much larger than other soft condensed matter fluctuation induced forces such as those due to critical fluctuations \cite{Gambassi_Dietrich_2011,Gambassi_2009} or Goldstone modes \cite{Kardar_Golestanain_1999} and, unlike critical fluctuation induced forces, they generically exist in the entire phase diagram. The NE results on the average fluctuation induced pressure has motivated us to study other aspects of this pressure, such as work and the fluctuating work. In particular, the aim of this paper is to understand aspects of the work distribution in a NESS in a systems where long ranged spatial correlations are important. 

The study of the work distribution in a NESS has a substantial history \cite{Sasa_Tasaki_2006, Komatsu_et_al_2008}, \cite{Komatsu_et_al_2011}, \cite{Komatsu_2010}. For example, NESS versions of the JFT have been derived \cite{Nakagawa_2012, Komatsu_et_al_2015}. Given the fluctuation properties of the  JFT between two equilibrium states, the utility of these results are not obvious, especially for macroscopic systems (see below Eq.(1.1)). Most of the previous work was also very formal, and the examples considered were not rich enough to include the NE long-ranged correlations, and the mode-coupling terms that generate them. The physical way to proceed is to identify all soft, or conserved, variables denoted by, say, $\{A\}$,  and construct a dynamical distribution function in terms of both $\{A\}$ and the non-equilibrium parameters. In this way it is possible to consistently describe both the long-range correlations and the non-localities that generically exist in a NESS.  For the case of a constant temperature gradient, this is approximately done in Appendix B.

The main new results of this paper are as follows: (i) We introduce and compute the Casimir work in a NESS. (ii) We explicitly compute the work fluctuations in a NESS. These 'process' correlation functions are novel because they involve temperature correlations between systems of different size. The results are anomalous due to the long-ranged correlations that generally exist in a NESS. (iii) We examine fluctuations in the Jarzynski fluctuation theorem in systems with short-range correlations and conclude that they generally diverge exponentially fast as the system size increases. (iv) We explicitly compute the non-equilibrium  work distribution for a system with long-range correlations and show that it is very different from the work distribution in systems with short-range correlations. (v) We propose a new way to experimentally measure the Casimir force or pressure in a NESS.

The content of this paper is as follows. In Section II we discuss NE long-range correlations in a simple fluid with a constant temperature gradient. A simple NE thermodynamic process and a fluctuating work is defined. In Section III  we compute the averaged work, the work fluctuations, and the work distribution function. In Section IV we summarize the results of this paper and raise a number of other issues. In Appendix A we discuss some properties of a model work distribution where there are no long-range NE correlations, and in Appendix B we relate our starting expression for the fluctuating pressure to the non-equilibrium distribution function approach. Some technical details about the calculation of the work fluctuations are given in Appendix C.

\section{NE PRESSURE AND NE WORK FLUCTUATIONS}
\label{sec:II}

\subsection{Non-equilibrium temperature and pressure fluctuations}
\label{subsec:II.A}

To be specific, we consider a fluid with a temperature gradient in the $z$-direction. The dimension of the system in the $z$-direction is $L$ while in the perpendicular direction it is $L_{x}=L_{y}=L_{\perp}$ and we assume $L_{\perp}\gg L$. For most liquid systems the thermal conductivity varies little with temperature so we can assume a linear temperature profile given by,
\begin{equation}
T(z)=T_{0}+{\frac{\Delta T}{L}}z.
\label{eq:2.1}
\end{equation}
Here $\Delta T$ is the temperature difference between the two walls in the $z$-direction.

The longest ranged, and most important, fluctuations or correlations will all be related to the temperature fluctuations, $\delta T$, or entropy fluctuations, $\delta S$. At constant pressure they are related by $\delta S=\frac{c_p}{T} \delta T$. Here $\delta T$ is the fluctuating total temperature minus its average value given by Eq.(2.1). We assume periodic boundary conditions in the transverse direction and perfectly conducting walls at $z=0$ and $L$ so that as a function of position $\delta T(\mathbf{x})$ exactly vanishes at the walls. Below we will also assume stress-free boundary conditions on the velocity fluctuations. The Fourier representation of $\delta T(\mathbf{x})$ is,

\begin{equation}
\delta T(\mathbf{x})=\frac{2}{L}\sum_{N=1}\int_{\mathbf{k}_{\perp}}e^{i\mathbf{k_{\perp}}\cdot\mathbf{x}_{\perp}}\sin({\frac{N\pi z}{L}})\delta T(\mathbf{k}),
\end{equation}
where $\mathbf{k}$ is the wave-vector of the fluctuation and where $k_z=N\pi /L$ and $\mathbf{x}_{\perp}$  and $\mathbf{k}_{\perp}$ are the position and wave-vector perpendicular to the temperature gradient

The well-known expression for the small wavenumber ($k\sigma\ll 1$, with $\sigma$ the molecular diameter) behavior
of the temperature fluctuations \cite{Dorfman_Kirkpatrick_Sengers_1994, Kirkpatrick_Cohen_Dorfman_1982A,DeZarate_Sengers_2006} is,
\begin{eqnarray}
\left\langle \left|\delta T({\bf k})\right|^{2}\right\rangle _{\mathrm{NESS}}=  \frac{LL_{\perp}^2k_{B}T}{\rho D_{T}(\nu+D_{T})}\frac{(k_{\perp}\nabla T)^{2}}{k^{6}}.\
\end{eqnarray}
Here $\rho$, $\nu$ and $D_{T}$ are the mass density, the kinematic
viscosity, thermal diffusivity of the fluid, $k_{\perp}=\sqrt{k_x^2+k_y^2}$, and $k^2=k_z^2+k_{\perp}^2$.  All of the thermo-physical parameters in Eq.(2.3) may be identified with their spatially averaged values \cite{Kirkpatrick_DeZarate_Sengers_2016a}. We note that this correlation function is long ranged
as indicated by its $k^{-4}$ behavior at small wave numbers, while
the equilibrium temperature fluctuations are of very short range in space
with no singular behavior of the corresponding Fourier transforms
at small wave numbers. In real space the non-equilibrium temperature correlations scale with $L$, the system size \cite{DeZarate_Sengers_2006}.

In addition to the non-equilibrium temperature fluctuations discussed above, there are also equilibrium temperature fluctuations. For reasons discussed below Eq.(2.3), the effects of the equilibrium fluctuations are orders of magnitude smaller  than those typically associated with Eq.(2.3) \cite{DeZarate_Sengers_2006}. A qualitative estimate of the relative size of these two effects will be given below Eq.(3.12).

The long-range part of the NE pressure, and pressure fluctuations, can be expressed in terms of the fluctuating temperature. One can use either a non-linear fluctuating hydrodynamic argument, or use a mode-coupling approximation as is discussed in Appendix B. In either case one obtains \cite{Kirkpatrick_Dorfman_2015, Kirkpatrick_DeZarate_Sengers_2013, Kirkpatrick_DeZarate_Sengers_2014},

\begin{equation}
\tilde{p}_{\mathrm{NE}}(\mathbf{x})=A[\delta T(\mathbf{x})]^{2},
\end{equation}
where the tilde denotes a fluctuating quantity and the NE denotes that in averaging this quantity only NE contributions are to be included. Here,

\begin{eqnarray}
 A=\frac{\rho c_{p}(\gamma-1)}{2T}\left[1-\frac{1}{\alpha c_{p}}\left(\frac{\partial c_{p}}{\partial T}\right)_{p}+\frac{1}{\alpha^{2}}\left(\frac{\partial\alpha}{\partial T}\right)_{p}\right],
\end{eqnarray}
where $c_{p},\gamma,\alpha$ are, respectively, the specific heat capacity at
constant pressure, the ratio of specific heat capacities, and the coefficient
of thermal expansion.

With Eqs.(2.2) and (2.4), the average NE pressure, $\langle\tilde{p}_{\mathrm{NE}}(\mathbf{x})\rangle$ can be computed \cite{Kirkpatrick_DeZarate_Sengers_2013, Kirkpatrick_DeZarate_Sengers_2014},
\begin{equation}
\langle\tilde{p}_{\mathrm{NE}}(\mathbf{x})\rangle=p_{\mathrm{NE}}(z)=A\langle[\delta T(\mathbf{x})]^{2}\rangle,
\end{equation}
with $p_{\mathrm{NE}}(z)$ the $z$-dependent NE pressure. From now on, all angular brackets in the main text denote a NESS average. Mechanical equilibrium requires that the spatial derivative of the total pressure vanish. This constraint together with particle number conservation leads \cite{Kirkpatrick_DeZarate_Sengers_2016a,Kirkpatrick_DeZarate_Sengers_2016b} to a $L$-dependent total pressure given by,
\begin{equation}
p(L)=p_{\mathrm{eq}}+{\overline p}_{\mathrm{NE}}(L).
\end{equation}
where $p_{eq}$ is the spatially averaged equilibrium, or local-equilibrium, pressure and ${\overline p}_{\mathrm{NE}}$ is the spatially averaged NE pressure,
\begin{equation}
{\overline p}_{\mathrm{NE}}(L)=\frac{1}{L}\int_{0}^{L}p_{\mathrm{NE}}(z)dz.
\end{equation}
Explicitly one obtains,
\begin{equation}
\overline{p}_{{\mathrm{NE}}}(L)=\frac{k_{{\rm B}}{T}^{3}}{48\rho\pi D_{T}\left(\nu+D_{T}\right)}\ AL\left(\frac{\nabla T}{{T}}\right)^{2}.
\end{equation}
In analogy to the Casimir pressure or force in equilibrium fluids near their critical points, we have called this the NE Casimir pressure or force because it is also due to long-range correlation effects.

Using Eqs.(2.4) and (2.3) we can also compute the pressure fluctuations. From Eq.(2.4), $\delta\tilde{p}_{\mathrm{NE}}(\mathbf{x})=A[(\delta T(\mathbf{x}))^{2}-\langle(\delta T(\mathbf{x}))^{2}\rangle]$.
We will assume in general that the NE temperature fluctuations are Gaussian distributed \footnote{This is consistent with using fluctuating hydrodynamics linearized about the steady state to obtain Eq.(2.3). The Gaussian nature of the temperature fluctuations then follows from the assumed Gaussian noise distribution.}. With this, we find the non-equilibrium pressure correlation function is

\begin{equation}
<\delta\tilde{p}_{\mathrm{NE}}(\mathbf{x})\delta\tilde{p}_{\mathrm{NE}}(\mathbf{y})>_{\mathrm{cumulant}}
=2A^{2}[<\delta T(\mathbf{x})\delta T(\mathbf{y})>]^{2}.
\end{equation}

At fixed temperature gradient, this correlation scales as $L^2$, because the temperature fluctuation scales as $L$. Eq.(2.10) will be used to compute the work fluctuations in Section III.B.

\subsection{Non-Equilibrium Casimir work}
\label{subsec:II.B}

The spatially averaged NE fluctuating pressure is,

\begin{equation}
{\tilde p}_{\mathrm{NE}}(L)=\frac{1}{LL_{\perp}^2}\int_{0}^{L}dz\int d{\mathbf{x}}_{\perp}\tilde{p}_{\mathrm{NE}}(\mathbf{x}),
\end{equation}

where we have explicitly indicated only the $L$-dependence.

If the system expands in the $z$-direction from length $L$ to length $L(1+\Delta)$, then the fluctuating non-equilibrium work is,

\begin{equation}
{\tilde{W}}_{\mathrm{NE}}(L\rightarrow L(1+\Delta))=-{L_{\perp}^2}\int_{L}^{L(1+\Delta)}dL_{1}{\tilde p}_{\mathrm{NE}}(L_{1}).
\end{equation}
To carry out the integral in Eq.(2.12) the thermodynamic path needs to be specified in more detail. This is done in the next subsection.

This work can be called the non-equilibrium Casimir work because the contribution  to ${\tilde p}_{\mathrm{NE}}(L)$ we consider is due to long-range correlations.

\subsection{Thermodynamic process}
\label{subsec:II.C}

To compute $\tilde{W}_{\mathrm{NE}}$ we need to determine how density, temperature, and other quantities change with $L$. That is, we need to specify a particular path or protocol in thermodynamic space. We will choose one that is computationally simple. In particular, we choose to use a protocol where the system size in the $z$-direction is quasi-statically changed to $L(1+\Delta )$. We imagine a source of particles so that the overall number density is fixed. We also fix the temperature difference, $\Delta T$, between the boundaries in the $z$-direction. With this the spatially averaged temperature and density are unchanged.

The equilibrium pressure is then independent of $L$ so that the equilibrium work is,
\begin{equation}
W_{\mathrm{eq}}=-p_{\mathrm{eq}}{L_{\perp}}^{2}L\Delta.
\end{equation}
For use later on, we define the total average work and the average NE work as,
\begin{equation}
\begin{split}
\overline{W}=W_{\mathrm{eq}}+\langle {\tilde{W}}_{\mathrm{NE}}(L\rightarrow L(1+\Delta))\rangle \\ \equiv W_{\mathrm{eq}}+
{\overline{W}}_{\mathrm{NE}}.
\end{split}
\end{equation}
Because it is due to a long-range fluctuation effects, ${\overline{W}}_{\mathrm{NE}}$ will be called the average Casimir work.

In general we can now compute $\tilde{W}_{\mathrm{NE}}$, given by Eq.(2.12), with the only $L$-dependences coming from  $\nabla T={\Delta T}/L$ and from the Fourier components in Eq.(2.2). 

The short-range work fluctuations that are associated with the equilibrium or local-equlibribm pressure will not be explicitly considered here. A general discussion of the work distribution for systems with short-range correlations is given in Appendix A.

\section{THE CALCULATIONS}
\label{sec:III}

\subsection{The average work}
\label{subsec:III.A}

With Eqs.(2.9), (2.4), (2.2), and (2.3) the average NE work is,

\begin{equation}
\langle {\tilde{W}}_{\mathrm{NE}}(L\rightarrow L(1+\Delta))\rangle={\overline{W}}_{\mathrm{NE}}=-C{L_{\perp}^2}ln (1+\Delta),
\end{equation}
where for use later on, we have defined the amplitude,
\begin{equation}
C=\frac{k_{B}TA(\Delta T)^{2}}{48\pi \rho D_{T}(\nu+D_{T})}.
\end{equation}

In the absence of long-range NE correlations or mode-coupling effects there is an additional contribution to the NE work that follows from including higher order gradient terms in a hydrodynamic description. If one calculates the diagonal part of the stress tensor in this generalized, or Burnett, hydrodynamic description one obtains a NE contribution to the pressure given by \cite{McLennan_1973, Wong_et_al_1978},

\begin{equation}
p_{\mathrm{NE}}=\kappa_{\mathrm{NL}}(\nabla T)^2=\kappa_{\mathrm{NL}}(\frac{\Delta T}{L})^2.
\end{equation}
Here $\kappa_{\mathrm{NL}}$ is a kinetic coefficient commonly referred to as a Burnett coefficient. With the thermodynamic protocol used above, the corresponding NE Burnett (NEB) work is,

\begin{equation}
W_{\mathrm{NEB}}((L\rightarrow L(1+\Delta))=-\kappa_{\mathrm{NL}}{L_{\perp}^2}\frac{(\Delta T)^2}{L}\frac{\Delta}{(1+\Delta)}.
\end{equation}
That is, this contribution is of $O(1/L)$ and sub-leading compared to the one from long-range non-equilibrium one, Eq.(3.1). 

To compare these two contributions we use that a natural length that appears in $C$ in Eq.(3.1) is,
\begin{equation}
l=\frac{k_{B}T}{\rho D_{T}(\nu+D_{T})}.
\end{equation}
We note that for water at STP, $l\approx3\times10^{-9}\mathrm{cm}$. If we take the ratio of Eqs.(3.1) and (3.4), we obtain in the small $\Delta$ limit,
\begin{equation}
\frac{W_{\mathrm{NEB}}((L\rightarrow L(1+\Delta))}{\langle {\tilde{W}}_{\mathrm{NE}}(L\rightarrow L(1+\Delta))\rangle}\propto\frac{{\ell}^2}{lL}\leq\frac{\ell}{L}.
\end{equation}
Here $\ell$ is the mean-free path and we have used that for liquid state densities $l\geq\ell$. Because for liquid state densities $\ell$ is a fraction of a molecular diameter, we conclude that the for all reasonable cases the average ${\tilde{W}}_{\mathrm{NE}}$ is orders of magnitude larger than $W_{\mathrm{NEB}}$.

A way to measure the Casimir work, ${\overline{W}}_{\mathrm{NE}}$, via a heat measurement will be discussed in Section IV.

%An experimental measurement of ${\overline{W}}_{\mathrm{NE}}=-Q_{\mathrm{NE}}$ via a heat measurement for the thermodynamic protocol used here would be a novel way to verify or not, the predicted \cite{Kirkpatrick_DeZarate_Sengers_2013, Kirkpatrick_DeZarate_Sengers_2014} NE Casimir pressure for this NESS. Here ......

\subsection{Work fluctuations}
\label{subsec:III.B}

The NE work fluctuations are,
\begin{equation}
\langle (\delta{\tilde W}_{\mathrm{NE}})^2\rangle={L_{\perp}^4}\int_{L}^{L(1+\Delta)}dL_{1}dL_{2}\langle \delta{\tilde p}_{\mathrm{NE}}(L_1) \delta{\tilde p}_{\mathrm{NE}}(L_2)\rangle.
\end{equation}
The equilibrium work fluctuations are sub-leading compared to Eq.(3.7) (see discussion remarks 2 and 3 in Section IV). Using Eqs.(2.9),(2.4), and (2.10), and a Gaussian approximation on the resulting correlation functions, gives,

\begin{equation}
\begin{split}
\langle (\delta{\tilde W}_{\mathrm{NE}})^2\rangle= 2A^2{L_{\perp}^4} \\ [\prod_{i=1}^{2}\int_{L}^{L(1+\Delta)}dL_{i}\int_{0}^{L_{i}}\frac{d{\mathbf{x}_{i}}}{L_{i}L_{\perp}^2}]\langle\delta T({\mathbf{x}}_1)\delta T({\mathbf{x}}_2)\rangle^{2}.
\end{split}
\end{equation}
The temperature correlation functions in Eq.(3.8) are more complicated than they naively appear because the temperature fluctuations are in two systems of different size.

The work fluctuations can be calculated by using Eqs.(2.2), (2.3) and (3.2). The result can be written,
\begin{equation}
\langle (\delta{\tilde W}_{\mathrm{NE}})^2\rangle={\frac{64\pi C^{2}}{1575}}{L^2}{L_{\perp}^2}G(\Delta).
\end{equation}
Here we have defined a dimensionless function of $\Delta$, $G(\Delta)$. The calculation is somewhat tedious and some technical details are given in Appendix C. We obtain,
\begin{equation}
G(\Delta)={\frac{\Delta^2}{(1+\Delta)^3}}[1+\frac{8}{3}\Delta+3\Delta^2+\frac{8}{5}\Delta^3+\frac{\Delta^4}{3}].
\end{equation}
For future use we note that $G$ has the limiting values,
\begin{equation}
G(\Delta\rightarrow 0)={\Delta}^2,
\end{equation}
and,
\begin{equation}
G(\Delta\gg 1)\approx\frac{{\Delta}^3}{3}.
\end{equation}

Finally, if we compare the non-equilibrium work fluctuations to equilibrium  work fluctuations, then the ratio is $O((\frac{\Delta T}{T})^4\frac{L}{\ell})$ \cite{Kirkpatrick_Dorfman_2015}. Further, if we consider mixed terms with both equilibrium and non-equilibrium fluctuations they are even less important.Therefore for moderate system sizes, and for moderate temperature differences, the non-equilibrium  work fluctuations dwarf the equilibrium ones.

\subsection{The non-equilibrium work distribution function}
\label{subsec:III.C}

We next determine the non-equilibrium work distribution function. For notational simplicity we denote the NE work distribution function, $\rho({\tilde W}_{\mathrm{NE}})$, by ${\rho_{\mathrm{NE}}(W)}$, with $W$ the fluctuating work in this sub-section. 

As a first step we determine the length and $\Delta T$ scaling properties of the higher-order cumulants. To do this we note that diagrammatically at one-loop order, the average pressure, Eq.(2.6), is related to a simple loop, the pressure fluctuations, Eq.(2.10), are a football (a digon), the third-order cumulant will be a triangle, and so on.  Because $\left\langle \left|\delta T({\bf k})\right|^{2}\right\rangle\propto{(\nabla T)^{2}k^{-4}}$, in accordance with Eq.(2.3), this means at every order there will be an extra factor of $(\nabla T)^{2}L^{4}\propto (\Delta{T}L)^2$. There is also a overall factor of ${L_{\perp}^{2}}/{L^2}$. This leads to the scaling
\begin{equation}
\langle (\delta{\tilde W}_{\mathrm{NE}})^n\rangle_{\mathrm{cumulant}}\propto (-\Delta)^{n}\frac{{L_{\perp}^{2}}}{L^2}\Big(L\Delta{T}\Big)^{2n}.
\end{equation}

Next the factors in Eq.(3.13) need to be determined. First we specialize to the case where $\Delta\ll 1$. The $n$-th cumulant can then be explicitly determined by counting the number of ways that $n$-$(\delta T)^2$ factors can be correlated in a cumulant correlation, and by generalizing Eqs.(C6) and (C7) to general $n$. After a straight-forward but lengthy calculation we obtain,
\begin{equation}
\begin{split}
\langle (\delta{\tilde W}_{\mathrm{NE}})^n\rangle_{\mathrm{cumulant}}\equiv \kappa_{n}=\frac{\pi {L_{\perp}}^2}{8L^2}\zeta(4n-2) \\ \frac{n!(n-1)!(2n-2)!}{(3n-1)!}\Big(\frac{-96\Delta{\overline{C}}L^{2}(\Delta{T})^{2}}{{\pi}^{3}}\Big)^{n},
\end{split}
\end{equation}
where $\overline{C}=C/(\Delta{T})^{2}$ and $\zeta(4n-2)$ is a Riemann zeta function. Equation (3.14) defines the $n$-cumulant, $\kappa_{n}$.

With Eq.(3.14) we can determine $\rho_{\mathrm{NE}}(W)$ as follows. First  we define a cumulant generating function, $K(t)$ by,
\begin{equation}
\begin{split}
K(t)=\ln \Big(\int{dWe^{Wt}\rho_{\mathrm{NE}}(W)})\Big) =\sum_{n=1}\frac{\kappa_{n}t^n}{n!},
\end{split}
\end{equation}
The work distribution is now formally given by the inverse transform,
\begin{equation}
\rho_{\mathrm{NE}}(W)=\int dte^{-Wt+K(t)}.
\end{equation}
The integral in Eq.(3.16) can be evaluated using saddle-point methods because the scale of $W$ grows with system size. The important feature in the evaluation of Eq.(3.16) is the convergence property, or singularity structure, of $K(t)$, Eq.(3.15), which in turn is determined by the large $n$-behavior of $\kappa_n$. In this limit, ${\kappa_n}/n!\approx a(-b{\Delta}/|\Delta|)^{n}/n^{3/2}$, with
\begin{equation}
a=\frac{\pi^{3/2}\sqrt{3}}{16}\frac{{L_{\perp}^2}}{L^2}
\end{equation}
an aspect ratio measure, and
\begin{equation}
b=\frac{128|\Delta|{\overline{C}}}{9{\pi}^3}(L\Delta T)^2\propto |\Delta|\Big(\frac{\Delta TL}{Tl}\Big)^2n_{\mathrm{av}}l^3.
\end{equation}
closely related to the average non-equilibrium work. $n_{\mathrm{av}}$ in Eq.(3.18), and below, is the average number density and is it is what makes the various factors dimensionally correct.

If we use all of this we see that $K(t)$ only converges for $|b|t<1$, and has a square root singularity at $|b|t=1$. This singularity in turn determines the saddle-point in Eq.(3.16) and from that $\rho_{\mathrm{NE}}(W)$ can be determined for large ${L_{\perp}}^2/L^2$ (cf. below Eq.(3.19)). Neglecting non-exponential pre-factors we obtain,
\begin{equation}
\rho_{\mathrm{NE}}(W)\propto\exp\Big[-\frac{|W|}{b}-\frac{\pi a^{2}b}{|W|}\Big]\theta{(-W\Delta)}.
\end{equation}

Consistent with Section III.B, Eq.(3.19) gives $|\langle W\rangle|=\sqrt{\pi}ab\propto {L_{\perp}^2}(\Delta T)^2$ \footnote{The coefficient in this expression for $|\langle W\rangle|$ is not exactly that given by Eq.(3.1) because in doing the cumulant sum to obtain Eq.(3.19) we made approximations that are only qualitatively correct for small $n$. For $n\geq 2$ the large $n$ approximation is quantitatively accurate.}. With this the natural scale of $W$ is $ab$ so that the scale of the exponential in Eq.(3.19) is $\propto a\propto {L_{\perp}^2}/L^2$. This is to be contrasted with the scale of $W$ in Eq.(A2), which is $N\propto L^3$. Physically this implies that the $W$ distribution in a NESS with long-range correlations is extremely broad compared to the work distribution in a system with short-range correlations. In the latter case, it scales like $N$ due to the central limit theorem.

To compare and contrast this distribution function for a system with long-range correlations with one with only short-range correlations in more detail, we compute the relative fluctuation measure $\epsilon_{\Omega}$ defined by Eq.(1.2) and in Appendix A. Setting $\beta =1$ and again using saddle-point methods we obtain,
\begin{equation}
\epsilon_{\Omega}\approx e^{F(a,b)},
\end{equation}
with
\begin{equation}
F(a,b)=2\sqrt{\pi}a\Big(2\sqrt{1+b}-1-\sqrt{1+2b}\Big).
\end{equation}
Here we have taken the case where $\Delta<0$ so that the work distribution is non-zero only for $W>0$ and averages like $\langle e^{-W}\rangle$ and $\langle e^{-2W}\rangle$ exist for all $b$. 

The two interesting limits are,
\begin{equation}
F(a,b\gg 1)\approx 4(1-\frac{1}{\sqrt{2}})a\sqrt{b\pi}\propto \frac{{L_{\perp}^2}}{Ll}\frac{\Delta T}{T}\sqrt{|\Delta|n_{\mathrm{av}}l^3}
\end{equation}
and
\begin{equation}
F(a,b\ll 1)\approx \frac{\sqrt{\pi}}{2}ab^2\propto \frac{L^{2}{L_{\perp}}^{2}}{l^4}\Big(\frac{\Delta T}{T}\Big)^4\Delta^2(n_{\mathrm{av}}l^3)^2 .
\end{equation}
In typical macroscopic experiments, Eq.(3.22) holds so that for fixed $\Delta T$ the $\Omega$ fluctuations defined by Eq.(1.2), and in Appendix A, are dramatically different than in the simple model with only short-range correlations that is considered in Appendix A, where we show that the analogous result scales with total system size, i.e., it is extensive. The size scaling of $F$ in Eq.(3.22) is non-extensive. On the other hand, Eq.(3.23) indicates that for small or mesoscopic systems, the size scaling of $F$ is super-extensive for fixed $\Delta T$, indicating an interesting crossover between the two limits.

Finally, we note  that the Gaussian approximation to Eq.(3.19) is,
\begin{equation}
\rho_{\mathrm{NE}}(W)\propto\exp\Big[-\frac{(\delta W)^{2}}{\sqrt{\pi}ab^{2}}\Big].
\end{equation}
If this is used to compute $F(a,b)$, one obtains Eq.(3.23) which indicates that the Gaussian approximation is valid only for small $b$.

\section{DISCUSSION}
\label{sec:IV} 

In this paper we have calculated the average work, the work fluctuations, and the work distribution function for a fluid in a uniform temperature gradient. The long-range correlations in this system make these quantities and functions dramatically different from what would be expected based on equilibrium considerations where correlations are generically of short-range. In particular, the NE work is a factor of $L$ larger than any NE work in the absence of long-range NE fluctuations, and the work distribution is very broadly distributed compared to an equilibrium work distribution. This last point leads to anomalous length scaling of the work fluctuations. It also dramatically modifies the size scaling of fluctuations that are important in determining the utility of the Jarzynski fluctuation theorem.

All of these features indicate just how different NE systems are from equilibrium ones.

We conclude with a number of other remarks:

\enumerate

\item We have used Eq.(2.4) as our starting point for computing pressure and work and their fluctuations. A natural question is the importance of higher-order terms, say those going like $A_{3}(\delta T({\mathbf{x}}))^{3}$ or $A_{4}(\delta T({\mathbf{x}}))^{4}$. If either of these terms are used to compute the pressure or pressure fluctuations it is easy to confirm that they are at most of $O({\overline p}_{\mathrm{NE}}(L)/p_{\mathrm{eq}})$ compared to contribution from Eq.(2.4). Since ${\overline p}_{\mathrm{NE}}(L)/p_{\mathrm{eq}}\propto l/L$, see Eq.(3.5), they are always small.

In transient non-equilibrium systems, the assumption of local equilibrium can break down, and the calculation of fluctuation-induced forces  is much more complicated \cite{Gambassi_2008, Dean_2012, Furukawa_et_al_2013}.
\item In equilibrium systems, without Goldstone modes, correlations are of short range, and typical root mean-squared fluctuations scale with volume $V$ as $\propto 1/{\sqrt{V}}$. In systems with long-ranged correlations this is not the case \cite{Kirkpatrick_Dorfman_2015}. For the NE work considered here, the appropriate measure is,
\begin{equation}
\epsilon_{\tilde{W}_{\mathrm{NE}}}=\frac {\langle{{\tilde W}_{\mathrm{NE}}}^2 \rangle -{\langle{\tilde{W}_{\mathrm{NE}}}\rangle}^{2}}{{\langle{\tilde{W}_{\mathrm{NE}}} \rangle}^{2}}.
\end{equation}
With Eqs.(3.1),(3.9), and (3.10), we find that this scales as,
\begin{equation}
\epsilon_{\tilde{W}_{NE}}\propto\frac{L^2}{L_{\perp}^2}.
\end{equation}
Typically one imagines that $L$ and $L_{\perp}$ scale in the same way. In this sense, the relative work fluctuations do not decrease with system size. This feature is due to two things. First, the average NE work for the thermodynamic protocol used here is of $O(L^0)$, which is small. Second, the long-range nature of the pressure fluctuations \cite{Kirkpatrick_Dorfman_2015} or work fluctuations causes the numerator to be large.

Note that Eq.(4,2) is basically the inverse of Eq.(3.22) just like the $N$ in Eq.(A7) is the inverse of the relative work fluctuations in a system with only short-range correlations. Physically this illustrates that the relative $\Omega$ fluctuations measure the exponential of the inverse relative work fluctuations, and that the broader the work distribution, the smaller the $\Omega$ fluctuations.
\item To obtain Eq.(4.2) we used $\langle\tilde{W}_{\mathrm{NE}}\rangle\propto{O(L^{0}L_{\perp}^{2}({\Delta T})^2)}$ and $\langle ({\delta{\tilde W}_{\mathrm{NE}}})^2\rangle\propto O((\Delta T)^{4}L^{2}{L_{\perp}^2})$. If in Eq.(4.1) we used $W_{\mathrm{eq}}\propto{O(L{L_{\perp}}^2)}$, see Eq.(2.13), in the denominator, we would conclude that work fluctuations scale like ${\epsilon}_{\tilde W}\propto (\Delta T)^{4}/{L_{\perp}^2}$, which are still large compared to the naive equilibrium statistical mechanics result ${\epsilon}_{\tilde W}\propto 1/(L{L_{\perp}^2})$. This is turn is related to the anomalous pressure fluctuations discussed in \cite{Kirkpatrick_Dorfman_2015}.

Thermodynamic processes where $\nabla T$ is fixed rather than $\Delta T$ are also possible. In this case the leading, in $L$, contribution to the average work will generally be $\langle\tilde{W}\rangle\propto O(L({\nabla T})^2)$, and the relative mean-squared fluctuation will generally scale as, $\epsilon_{\tilde{W}_{\mathrm{NE}}}\propto\frac{L^2}{L_{\perp}^2}$, just as in Eq.(4.2).
\item The long-range correlations that generically exist in NE statistical mechanics profoundly cause NE properties to be very different from equilibrium properties. Non-equilibrium quantities do
not have virial expansions \cite{Dorfman_Kirkpatrick_Sengers_1994, Dorfman_Cohen_1965}. A local expansion of the fluxes or currents
in terms of powers of the gradients is also not possible \cite{Dorfman_Kirkpatrick_Sengers_1994,Belitz_Kirkpatrick_Vojta_2005,Ernst_et_al_1978}.
Other techniques such as maximizing an entropy (for example, the so-called
max cal method \cite{Presse_et_al_2013}) to obtain a non-equilibrium
distribution function will not work, at least in
their most naive form. 

The dramatic difference between the $\Omega$-fluctuations calculated in Section III.D for a NESS with long-range correlations and those computed in Appendix A for a system with short-range correlations is another example of how different non-equilibrium really is. In this context, the dramatic difference is caused by the broadness \cite{Kirkpatrick_Dorfman_2015} of the distributions in systems with long-range correlations.

%In this context, it would be interesting to explicitly compute the $\Omega$-fluctuations defined in Appendix A for a NESS with long-ranged correlations. In the Appendix we argued that we generally expect the relative $\Omega$-fluctuations to diverge exponentially fast as the system size increases. It is possible that in a NESS with long-ranged correlations, these relative fluctuations will be very different.

% In particular the long-ranged NESS correlations implies that the work distribution associated with these fluctuations will be very broad %compared to the usual equilibrium distributions. This in turns suggest that the relative $\Omega$-fluctuations for this work be smaller.
\item In equilibrium systems near critical points, or in systems with Goldstone modes, the Casimir force or pressure is related to the volume dependence of the energy or free energy. In non-equilibrium this is in general not the case because of heat and possibly because of the distinction between reversible work, $W_{\mathrm{rev}}$ and irreversible work, $W_{\mathrm{irr}}$. The sign convention we use for the first law is $\Delta E=\overline{W}+Q$, with $\Delta E$ the change in average internal energy, $\overline W$ the total average work, and $Q$ the total heat. In the thermodynamic process considered here there is no change in the average temperature, and the internal energy only changes because of the volume change at fixed density \footnote{Strictly speaking, the energy will be a function of the local temperature, $T(z)$, Eq.(2.1), and not simply the average temperature  $\overline{T}$. The replacement of $T(z)$ by $\overline{T}$ is an approximation that is only valid when  fluctuations in the spatial average of $T(z)$ are small. These fluctuations are readily computed and one finds $\overline{{\delta{T(z)}}^{2}}=(\Delta{T})^{2}/12$. If we normalize this with the average temperature, these fluctuations are indeed quite small, especially since typically $\Delta{T}\ll\overline{T}$.}. With a quasi-static process we can then, to a good approximation, split $\Delta E$, $\overline{W}$, and $Q$ into their equilibrium ($\Delta T=0$), at average temperature $\overline T$, and non-equilibrium  ($\Delta T\neq 0$) at the same average temperature, contributions as,
\begin{equation}
\Delta E_{\mathrm{eq}}=\overline{W}_{\mathrm{eq}}+Q_{\mathrm{eq}}
\end{equation}
and, 
\begin{equation}
{\overline{W}}_{\mathrm{NE}}=-Q_{\mathrm{NE}}.
\end{equation}
That is, the NE average work is given by the additional heat that occurs when $\Delta T\neq 0$ between the two plate in the $z$-direction.  Note too, that the terms in Eq.(4.3) are all of $O(L{L_{\perp}}^{2}\Delta)$, while those in Eq.(4.4) are predicted to be of $O(L_{\perp})^{2}\ln{(1+\Delta)}$. This different in $L$, and $\Delta$, scaling can also be used to distinguish the equilibrium contributions from the non-equilibrium ones. In any case, the average NE work is not equal to the internal energy change.

\item The average NE work, Eq.(3.1), is a factor $L$ larger than the naively expected result, Eq.(3.4), due to the long-range correlations that exist in a fluid with a uniform temperature gradient.

An experimental measurement of ${\overline{W}}_{\mathrm{NE}}=-Q_{\mathrm{NE}}$ via a heat measurement for the thermodynamic protocol used here would be a novel way to verify or not, the predicted \cite{Kirkpatrick_DeZarate_Sengers_2013, Kirkpatrick_DeZarate_Sengers_2014} NE Casimir pressure for this NESS.
\item A natural question is, just how general are our results.  Can or should one imagine universality classes for NE fluctuations or correlations, similar to, for example, in static \cite{Wilson_Kogut_1974} or dynamic \cite{Hohenberg_Halperin_1977} critical phenomena? Physically the effects of most interest are determined by long wavelength effects so these are reasonable questions. The number of possible universality classes (UCs) would be quite large. They would depend on the equilibrium statics, for example if Goldstone modes exist, and on the dynamics via how many conserved variables there are, and if mode-coupling non-linearities are involved. The UCs would also depend on the particular NESS. For example, correlations are much stronger and longer range in the case of a fluid with a temperature gradient than in a fluid with a velocity gradient or shear \cite{Lutsko_Dufty_2002, Sengers_DeZarate_2010}. In principle the NE work properties might also depend on the particular thermodynamic path. It is apparent that the number of distinct UCs will be quite large.

\endenumerate 

\acknowledgments
Discussions with Chris Jarzynski and Dave Thirumalai are gratefully acknowledged. The research was supported by the U.S. National Science Foundation under Grant DMR-1401449. 

\appendix

\section{\bf{ FLUCTUATIONS IN $e^{-{\beta}W}$ IN SIMPLE MODELS}}
\label{app:A}

Since the mid-1990's, so-called fluctuation theorems for either the work or entropy production in some thermodynamic process have been greatly studied. One of the most well-known is the Jarzynski fluctuation theorem which relates the average of minus the exponential of the microscopic work, $W$, $\Omega=e^{-{\beta}W}$, to the free energy change, $\Delta F$, in going from one thermodynamic state to another. The general and exact relation is,

\begin{equation} 
\langle \Omega\rangle=e^{-\beta \Delta F}, 
\end{equation}
where the angular brackets denotes an average over the work distribution $\rho (W)$. 

Here we show, first in a simple model, and then argue more generally, that this equality is pathological in that the relative fluctuations of the quantity $\Omega$ diverges exponentially as the system size or number of particles increases. Physically this implies that Eq.(A1) is in general not a computationally useful equality unless the system size is small. On the other hand, equilibrium statistical mechanics would suggest that $\ln \Omega$ would have more normal fluctuations, with relative root-mean-squared  fluctuations of $O(1/\sqrt{V})$. In the main text we have shown that even this is not generally true, due to the generic long-range correlations that exist in non-equilibrium. 

In \cite{Crooks_Jarzynski_2007} a thermodynamic work process in an ideal gas was considered. Setting $\beta=1$, their work distribution (here we take $N-1\approx N$) is,
\begin{equation}
\rho(W)={\mathcal{N}}\big({\frac{W}{\alpha}}\big)^{dN/2}e^{-W/\alpha}\theta(\alpha W),
\end{equation}
where $\mathcal{N}$  is a normalization factor and $\alpha$ is a volume change factor (related to $-\Delta$ used in the main text).
Now, saddle point methods can be used to compute any quantity for large $N$, but the saddle point changes if you multiply by factors exponentially small or large in $N$ such as $\Omega=e^{-W}$. 
A simple calculation gives,
\begin{equation}
\langle W\rangle={\frac {dN}{2}}\alpha,
\end{equation}
\begin{equation}
\langle \Omega\rangle=e^{-{\frac {dN}{2}}ln(1+\alpha)}.
\end{equation}
The Jarzynski fluctuation theorem then gives the free energy change,
\begin{equation}
\Delta F={\frac {dN}{2}}\ln(1+\alpha).
\end{equation}

The relative fluctuations in $\Omega$,  as defined in Eq.(1.2), is, 
\begin{equation}
\epsilon_{\Omega}=\frac {\langle{\Omega}^2 \rangle -{\langle\Omega \rangle}^{2}}{{\langle\Omega \rangle}^{2}},
\end{equation}

Again, this can be computed by using saddle-point methods to obtain, ignoring non-exponential pre-factors,
\begin{equation}
\epsilon_{\Omega}=e^{{\frac {dN}{2}}{\ln(1+\frac {{\alpha}^2}{1+2\alpha})}}-1\approx e^{{\frac {dN}{2}}{\ln(1+\frac {{\alpha}^2}{1+2\alpha})}}.
\end{equation}

That is, the fluctuations are exponentially large in $N$. Computationally that means an exponentially large number of trajectories  are needed to obtain a convergent result for $\langle\Omega\rangle$. This is to be contrasted with the relative root-mean-squared fluctuations in $W$ which scale like $1/\sqrt N$. 

More generally, any Gaussian distribution with a variance scaling like $1/\sqrt N$ will have an exponentially divergent (as $N\rightarrow\infty$) relative fluctuation for any quantity such as $\Omega$, that is exponentially small in $N$. Even if the starting Gaussian distribution is not exact, this argument indicates that exponentially divergent $\Omega$ fluctuations are generic.
\bigskip

\section{\bf{CONNECTION WITH THE NON-EQUILIBRIUM DISTRIBUTION APPROACH}}
\label{app:B}

The general methods of non-equilibrium statistical mechanics can be related to the non-linear fluctuating hydrodynamic approach used in the main text.  In this Appendix we show that the two approaches are equivalent. In particular, we use the non-equilibrium $Gamma$-space distribution functions of MacLennan \cite{MacLennan_1961} and Zubarev \cite{Zubarev_1962, Zubarev_1974}, combined with the mode-coupling theories of Kadanoff and Swift \cite{Kadanoff_Swift_1968} and of Kawasaki \cite{Kawasaki_1970}, to derive some of the results of Section II.

For the case of a NESS with a temperature gradient the $Gamma$-space distribution function is \cite{MacLennan_1961, Zubarev_1962, Zubarev_1974}

\begin{equation}
\rho=\rho_{\mathrm{L}}\exp[\mathcal{S}],
\end{equation}

where $\mathcal{S}$ is the total entropy production factor,

\begin{equation}
\mathcal{S}=\int_{-\infty}^{0}e^{\epsilon t}\hat{\mathbf{J}}_{Q}(\mathbf{x},t)\star\nabla\beta(\mathbf{x})dt.
\end{equation}

where $\beta(\mathbf{x})=1/k_BT(\mathbf{x})$ is the inverse local temperature and $\star$ denotes a spatial integration. In Eq.(B1), $\rho_{\mathrm{L}}$ is gamma-space local equilibrium distribution function,
\begin{equation}
\rho_{\mathrm{L}}=Q_{l}^{-1}\exp[-F_{m}(\mathbf{x})\star P_{m}(\mathbf{x})],
\end{equation}

where $P_{m}$ is the set of microscopic conserved variables, $\{P_{m}\}=(H,\mathbf{g},n)$
and $F_{m}$ are the set of conjugate thermodynamic, or hydrodynamic
variables, $\{F_{m}\}=(\beta,\mathbf{0},-\beta\mu)$. Here $H$ is the microscopic Hamiltonian, $\mathbf{g}$ is the phase-space local momentum density, $n$ is the micrscopic number density, and $\mu$ is the chemical potential. In
 Eq.(B1) $\hat{\mathbf{J}}_{Q}(t)$ is the microscopic
projected heat current at time $t$, $\star$ denotes that repeated
space indices are to be integrated over, and $\epsilon=0^{+}$
is an infinite past convergence factor. 

$\hat{\mathbf{J}}_{Q}=P_{\perp}\mathbf{J}_{E}$ is by construction orthogonal to the conserved quantities, so it does not obviously decay via a slow hydrodynamic process. Here $P_{\perp}=1-P$, with $P$ a projection operator (onto the conserved or hydrodynamic variables) and $\mathbf{J}_E$ is the energy current. However, $\hat{\mathbf{J}}_{Q}$ does have a projection onto products of hydrodynamic modes, ${A_{\alpha\mathbf{k}}}$ which are also slow modes. To take this into account one introduces a product of hydrodynamic modes projector,

\begin{equation}
\mathcal{P}=\frac{1}{2}\sum_{{\mathbf{k}_{1}\mathbf{k}_{2}},\alpha,\beta}|[A_{\alpha{\mathbf{k}}_{1}}A_{\beta{\mathbf{k}}_{2}}]\rangle\langle [A_{\alpha{\mathbf{k}}_{1}}A_{\beta{\mathbf{k}}_{2}}]|.
\end{equation}
Here the product modes are constructed to be orthogonal to single modes so that,
\begin{equation}
[AA]=AA-PAA,
\end{equation}
where $P$ is the single mode projection operator.
In Eq.(B4) the modes $A$ are taken to be orthonormal \cite{Ernst_Hauge_vanLeeuwen_1976}, the factor of two ensures $\mathcal{P}^2=\mathcal{P}$, and $\langle |\rangle$ denotes a local-equlibribm average.

The heat current in Eq.(B2) is then written,
\begin{equation}
\hat{\mathbf{J}}_{Q}(t)=\mathbf{\mathcal{J}}_Q(t)+\mathcal{P}\hat{\mathbf{J}}_{Q}(t),
\end{equation}
where $\mathcal{J}_Q(t)$ is the heat current orthogonal to both hydrodynamic modes and products of hydrodynamic modes. The entropy production factor is written,
\begin{equation}
\mathcal{S}=\mathcal{S}_{\perp}+\mathcal{S}_2,
\end{equation}
with the part of $\mathcal{S}$ proportional to a product of modes given by,
\begin{equation}
\begin{split}
\mathcal{S}_2={\frac{1}{2}} \\ \sum_{{\mathbf{k}_{1}},\alpha,\beta}|[A_{\alpha{\mathbf{k}}_{1}}A_{\beta,{-\mathbf{k}}_{1}}]\rangle D_{\alpha\beta}({\mathbf{k}}_1).
\end{split}
\end{equation}
Here,

\begin{equation}
\begin{split}
D_{\alpha\beta}({\mathbf{k}}_1)\delta_{{{\mathbf{k}}_1+{\mathbf{k}}_2},\mathbf{0}}= \\ \langle A_{\alpha{\mathbf{k}}_{1}}A_{\beta{\mathbf{k}}_{2}}|\int_{-\infty}^{0}e^{\epsilon t}\hat{\mathbf{J}}_{Q}(\mathbf{x},t)\rangle\star\nabla\beta(\mathbf{x})dt.
\end{split}
\end{equation}
With $D_{\alpha\beta}({\mathbf{k}}_1)$ the one-loop NE contribution to the $\langle A_{\alpha{\mathbf{k}}_{1}}A_{\beta,-{\mathbf{k}}_{1}}\rangle_{\mathrm{NESS}}$ correlation function as can be seen by expanding Eq.(B1) to $O(\mathcal{S})$ \cite{Kirkpatrick_Cohen_Dorfman_1982A,Kirkpatrick_Cohen_Dorfman_1982B}.

For the NESS considered here the most important fluctuations are the entropy, or temperature at constant pressure, fluctuations. These fluctuations are defined in terms of conserved quantity fluctuations using thermodynamics \cite{Ernst_Hauge_vanLeeuwen_1976}. With this, Eq.(B8) is,
\begin{equation}
\begin{split}
\mathcal{S}_2= \frac{1}{2(k_{B}T)^2} \\ {\sum_{{\mathbf{k}_{1}}}|[\delta S_{{\mathbf{k}}_{1}}\delta S_{-{\mathbf{k}}_{1}}}]\rangle D_{TT}({\mathbf{k}}_1).
\end{split}
\end{equation}

The final connection with the fluctuating hydrodynamic approach results of Section II is made by using that $A$, given by Eq.(2.5), is also given by \cite{Ernst_Hauge_vanLeeuwen_1976},
\begin{equation}
A={\frac{1}{2V(k_{B}T)^2}}\langle\tilde  p_{\mathbf 0}|[\delta S_{\mathbf{0}}\delta S_{\mathbf{0}}]\rangle,
\end{equation}
where to avoid confusion we have denoted the fluctuating microscopic phase space pressure \cite{Ernst_Hauge_vanLeeuwen_1976} with a tilde.
If we now use the same projector, $\mathcal{P}$, we can write the two-mode contribution of the microscopic pressure as,
\begin{equation}
{\tilde{p}}_{\mathbf{k}}=\frac{1}{2(k_{B}c_p)^{2}}\sum_{{\mathbf{k}}_1, {\mathbf{k}}_{2}} |[\delta{S}_{{\mathbf{k}_{1}}}\delta{S}_{{\mathbf{k}_{2}}}]\rangle\langle[\delta{S}_{{\mathbf{k}_{1}}}\delta{S}_{{\mathbf{k}_{2}}}]|{\tilde{p}}_{\mathbf{k}}\rangle.
\end{equation}
For small wave-numbers, with Eq.(B11), this becomes,
\begin{equation}
{\tilde{p}}_{\mathbf{k}}=A\Big{(}\frac{T}{c_p}\Big{)}^{2}\sum_{{\mathbf{k}}_1} |[\delta{S}_{{\mathbf{k}_{1}}}\delta{S}_{{\mathbf{k}-\mathbf{k}_{1}}}]\rangle.
\end{equation}
In real space, in terms of temperature fluctuations, this is identical to Eq.(2.4).
%\begin{equation}
%\tilde{p}_{NE}(\mathbf{x})=A[\delta T(\mathbf{x})]^{2}.
%\end{equation}

To check for consistency, we expand Eq.(B1) to first order and use Eqs.(B10) and (B11) to compute $\langle\tilde{p}_{\mathrm{NE}}(\mathbf{x})\rangle$ and obtain \cite{Kirkpatrick_DeZarate_Sengers_2014},
\begin{equation}
\langle\tilde{p}_{\mathrm{NE}}(\mathbf{x})\rangle_{\mathrm{NESS}}=AD_{TT}(\mathbf{x=0}).
\end{equation}
That is,  Eq.(2.6), in terms of $D_{TT}$, which is the one-loop result for this quantity \cite{Kirkpatrick_Cohen_Dorfman_1982A,Kirkpatrick_Cohen_Dorfman_1982B}.

To compute terms of two and higher loop order, additional higher mode terms must be retained in this effective theory. In particular, it is not consistent to use Eq.(B10) in the exponential and then construct a self-consistent theory for $D_{TT}({\mathbf{k}}_1)$.

Finally we remark that the NE distribution approach used here cannot naturally be used to compute the pressure fluctuations needed in Eq.(3.7) for the work fluctuations since that involves pressure fluctuations in two systems of different size. The most straightforward way to compute the correlation function needed in Eqs.(3.7) and (3.8) is with fluctuating hydrodynamics. Note, however, Eq.(B13) is generally the two-mode contribution to the fluctuating pressure.

\bigskip

\section{\bf{WORK-FLUCTUATION RESULT}}
\label{app:C}

To obtain the function $G(\Delta)$, given by Eqs.(3.8) and (3.9), we use Eq.(2.2) to write,
\begin{equation}
\begin{split}
\langle\delta T(\mathbf{x}_{1})\delta T(\mathbf{x}_{2})\rangle=4 \\ [\prod_{i=1}^{2}\frac{1}{L_i}\sum_{N_{i}=1}\int_{\mathbf{k}_{i,\perp}}e^{i\mathbf{k}_{i,\perp}\cdot{\mathbf{x}_{i,\perp}}}\sin{(\frac{N_{i}\pi z_i}{L_i})}]  \\  \langle\delta T(\mathbf{k}_{1})\delta T(\mathbf{k}_{2})\rangle.
\end{split}
\end{equation}
We then insert the square of Eq.(C1) into Eq.(3.8) and carry out the spatial integrals to obtain,
\begin{equation}
\begin{split}
\langle (\delta{\tilde W}_{NE})^2\rangle= 8{A^2} \\ [\prod_{i=1}^{2}\int_{L}^{L(1+\Delta)}{\frac{dL_{i}}{L_{i}^2}}\sum_{N_{i}=1}\int_{\mathbf{k}_{i,\perp}}] \\ |\langle\delta T(\mathbf{k}_{1})\delta T(\mathbf{k}_{2})\rangle|^2.
\end{split}
\end{equation}

If we use, say, fluctuating hydrodynamics \cite{Ronis_Procaccia_1982, DeZarate_Sengers_2006} to compute the correlation functions in Eq.(C2) then integrals like,
\begin{equation}
\begin{split}
I=  \theta (L_{2}-L_{1}) \\ \int_{0}^{L_{1}}dz_{1}\sin{(\frac{N_{1}\pi z_1}{L_1})}\sin{(\frac{N_{2}\pi z_1}{L_2})}
\end{split}
\end{equation}
appear. In a single-size correlation function where $L_{1}=L_{2}=L$, the integral is $I=\frac{L}{2}\delta_{N_{1},N_{2}}$. For the two size case, for large $L$'s, Eq.(C3) is well approximated by,
\begin{equation}
I=\theta (L_{2}-L_{1}){\frac{L_1}{2}}\delta_{\frac{N_{1}}{L_{1}},{\frac{N_{2}}{L_{2}}}}.
\end{equation}
Here it is understood that the Kronecker delta function replaces, for example, $N_2$ by the integer closest to $N_{1}L_{2}/L_{1}$. The same approach implies that the $(\nabla T)^2$ in Eq.(2.3) gets replaced by $\frac{(\Delta T)^2}{L_{1}L_{2}}$ in the two-size case. Using all of this, we can write Eq.(C2) as,
\begin{equation}
\begin{split}
\langle (\delta{\tilde W}_{\mathrm{NE}})^2\rangle=2{D^2}{L_{\perp}^2}\int_{L}^{L(1+\Delta)}dL_{1}dL_{2}\sum_{N=1}\int_{\mathbf{k}_{\perp}} 
\\  {\frac{[L_{1}^{2}\theta(L_{2}-L_{1})+L_{2}^{2}\theta(L_{1}-L_{2})]}{L_{1}^{4}L_{2}^{4}}}{\frac{k_{\perp}^4}{[{\frac{N^{2}{\pi}^2}{L_{1}^2}}+k_{\perp}^2]^{6}}},
\end{split}
\end{equation}
where $D=48\pi C$ with $C$ given by Eq.(3.2). Finally, using
\begin{equation}
\zeta(6)=\sum_{N=1}\frac{1}{N^6}=\frac{{\pi}^6}{945}
\end{equation}
and
\begin{equation}
\int_{0}^\infty dy \frac{y^5}{[1+y^2]^6}=\frac{1}{60},
\end{equation}
we evaluate Eq.(C5) and obtain Eqs.(3.9) and (3.10).

\end{document}